\documentclass [twocolumn, aps, prl, superscriptaddress, showpacs] {revtex4}
\usepackage{graphicx,latexsym,makeidx}
\begin{document}
\draft
\title {Direct Measurement of the Fermi Energy in Graphene Using a Double Layer Structure}
\author {Seyoung Kim}
\affiliation {Microelectronics Research Center, The University of
Texas at Austin, Austin, TX 78758}
\author {Insun Jo}
\affiliation {Department of Physics, The University of Texas at
Austin, Austin, TX 78712}
\author {D. C. Dillen}
\affiliation {Microelectronics Research Center, The University of
Texas at Austin, Austin, TX 78758}
\author {D. A. Ferrer}
\affiliation {Microelectronics Research Center, The University of
Texas at Austin, Austin, TX 78758}
\author {B. Fallahazad}
\affiliation {Microelectronics Research Center, The University of
Texas at Austin, Austin, TX 78758}
\author{Z. Yao}
\affiliation {Department of Physics, The University of Texas at
Austin, Austin, TX 78712}
\author {S. K. Banerjee}
\affiliation {Microelectronics Research Center, The University of
Texas at Austin, Austin, TX 78758}
\author {E. Tutuc}
\affiliation {Microelectronics Research Center, The University of
Texas at Austin, Austin, TX 78758}
\date{\today}
\begin{abstract}
We describe a technique which allows a direct measurement of the relative Fermi energy in an electron system using a double layer structure,
where graphene is one of the two layers. We illustrate this method by probing the Fermi energy as a function of density
in a graphene monolayer, at zero and in high magnetic fields. This technique allows us to determine the Fermi velocity,
Landau level spacing, and Landau level broadening in graphene. We find that the $N=0$ Landau level broadening is larger by comparison to
the broadening of upper and lower Landau levels.
\end{abstract}
\maketitle

The Fermi energy is a fundamental property of an electron system, and thermodynamic measurements which probe the Fermi energy or density of states
are key to understanding the host material band structure and electron interaction effects. Although a number of thermodynamic properties,
such as specific heat \cite{gornik1985,wang1992}, magnetization \cite{jp1985}, magnetocapacitance \cite{smith}, or compressibility \cite{jp1994}
can directly probe the density of states in an electron system, accessing them experimentally becomes increasingly difficult at the micro- and nano-scale.
In the case of graphene \cite{geim}, magnetization and specific heat measurements are exceedingly difficult, and the accuracy of compressibility \cite{jp2010} and capacitance measurements \cite{ponomarenko,young,droscher} are also limited by the reduced sample dimensions. Using a double layer device structure where
graphene is one of the layers, we describe a technique which allows a direct measurement of the Fermi energy in an electron system with an accuracy which is
independent of the sample size. The underlying principle of the method discussed here is that an interlayer bias applied to bring the graphene layer
to the charge neutrality point is equal to the Fermi energy of the opposite electron system. We illustrate this technique by probing the Fermi energy
in a graphene layer, both at zero and in high magnetic fields. We show that this method allows an accurate determination of the Fermi
velocity in graphene, the Landau level spacing, and Landau level broadening in high magnetic fields.

Our samples are independently contacted graphene double layers, consisting of two graphene single layers separated by a thin dielectric as shown in Fig. 1(a) \cite{kim}. To fabricate such devices, we first mechanically exfoliate the bottom graphene layer from natural graphite onto a 280 nm thick SiO$_2$ dielectric,
thermally grown on a highly doped Si substrate. Standard e-beam lithography, Cr/Au deposition followed by lift-off, and O$_2$ plasma etching
are used to define a Hall bar device. A 4 to 7 nm top Al$_2$O$_3$ dielectric layer is deposited on the bottom layer by atomic layer deposition,
and using evaporated Al as a nucleation layer. The dielectric film thickness grown on graphene is further verified by
transmission electron microscopy in multiple samples. To fabricate the graphene top layer, a separate graphene single layer is mechanically
exfoliated on a SiO$_2$/Si substrate. After spin-coating polymetyl metacrylate (PMMA) on the top layer and curing, we etch the underlying
substrate with NaOH, and detach the top layer along with the alignment markers captured in the PMMA membrane. The membrane is transferred
onto the bottom layer device, and aligned. A Hall bar is subsequently defined on the top layer, completing the double layer graphene device.
Three samples were investigated in this study, all with similar results. We focus here on data collected from one sample with a 7.5 nm thick
interlayer dielectric, and with an interlayer resistance larger than 1 G$\Omega$. Both layer mobilities are 10,000 cm$^{2}$/V$\cdot$s.
Using small signal, low frequency lock-in techniques we probe the layer resistivities as a function of
back-gate bias ($V_{BG}$), and the inter-layer ($V_{TL}$) bias applied on the top layer. The bottom layer is maintained grounded during measurements.

\begin{figure}
\centering
\includegraphics[scale=0.27]{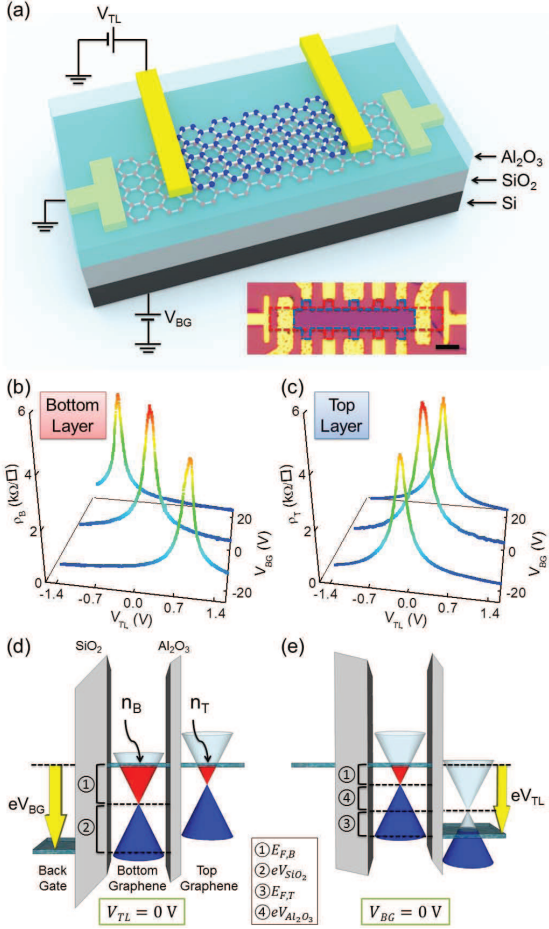}
\caption {\small{(color online) (a) Schematic representation of a graphene double layer separated by an Al$_2$O$_3$ dielectric,
and with a bottom SiO$_2$ dielectric. A back-gate ($V_{BG}$) and inter-layer ($V_{TL}$) bias can be applied on the Si substrate and top layer, respectively.
Lower right: optical micrograph of a graphene double layer device.  The red (blue) contour marks the bottom (top) layer. The scale bar is 5 $\mu$m.
(b,c) Layer resistivities measured as a function of $V_{TL}$ and $V_{BG}$ at $T=0.4$ K. (d,e) Band diagram of a graphene double
layer under an applied back-gate [panel (d)] or inter-layer [panel (e)] bias.}}
\end{figure}

Figure 1(b,c) data show the longitudinal resistivity of the bottom ($\rho_{B}$) and top ($\rho_{T}$) layer measured as a function of
$V_{TL}$, and at different $V_{BG}$ values \cite{gate_bias}. The data $\rho_{B,T}$ vs. $V_{TL}$ exhibit the ambipolar behavior characteristic of graphene,
and with a charge neutrality point which is $V_{BG}$-dependent. The shift of the charge neutrality point of the bottom layer as a function of
$V_{BG}$ is explained by picturing the bottom layer as a dual-gated graphene single layer, with the Si substrate as back-gate and the top graphene
layer serving as top-gate. The dependence of the $\rho_{T}$ vs. $V_{TL}$ data on $V_{BG}$ is more subtle, and implies an incomplete screening
by the bottom layer of the back-gate induced electric field. 

We can quantitatively explain the top and bottom layer density dependence on $V_{BG}$ and $V_{TL}$ using a simple band diagram model. Figure 1(d) shows the band diagram of the graphene bilayer at a finite $V_{BG}$, and with both layers at ground potential. For simplicity the back-gate Fermi energy and the two graphene layers charge neutrality points are assumed to be at the same energy at $V_{BG}=0$ V. Once a finite $V_{BG}$ is applied, charge densities are induced in both bottom ($n_{B}$), and top ($n_{T}$) layers. Consequently electric fields are built-in across both bottom SiO$_2$ and inter-layer Al$_2$O$_3$ dielectric. The applied $V_{BG}$
potential is the sum of the potential drop across the SiO$_2$ dielectric and the Fermi energy of the bottom layer:

\begin{equation}
eV_{BG}=e^{2}(n_B+n_T)/C_{SiO_2}+E_F(n_B)
\end{equation}

Here, $E_{F}(n_B)$ represents the Fermi energy of the bottom layer corresponding to a charge density $n_B$, and measured from the charge neutrality point;
$n_{B,T}$ and $E_F(n)$ are positive when the carriers are electrons, and negative when the carriers are holes. $C_{SiO_2}$ denotes the bottom dielectric capacitance per unit area.

Figure 1(e) shows a similar band diagram of the graphene bilayer, but in the presence of a finite inter-layer bias and at $V_{BG}=0$ V. Similarly to Fig. 1(d), the
applied $V_{TL}$ bias can be written as the sum of the potential drop across the Al$_2$O$_3$ dielectric and the Fermi energies of the two layers:

\begin{equation}
eV_{TL}=E_F(n_B)-(E_F(n_T)+e^{2}n_T/C_{Al_2O_3})
\end{equation}

Here, $E_F(n_T)$ represents the Fermi energy of the top layer at a charge density $n_T$, and $C_{Al_2O_3}$ is the interlayer dielectric capacitance per unit area.
In Fig. 1(e) the $V_{TL}$ bias is assumed to be positive, resulting in electrons (holes) induced in the bottom (top) layer. Although we derived Eqs. (1) and (2)
assuming $V_{TL}=0$ V, and $V_{BG}=0$ V respectively, the two equations hold at all $V_{BG}$ and $V_{TL}$ values. Most importantly,
we do not make any assumption with regard to the $E_F$ dependence on $n_B$ and $n_T$. As we show below, this dependence will be determined experimentally.

Figure 2(a) data show contour graphs of $\rho_{B}$ (left panel) and $\rho_{T}$ (right panel) as a function of $V_{BG}$ and $V_{TL}$.
The bottom layer resistivity dependence on gate bias is very similar to a dual-gated graphene monolayer \cite{kim},
showing an almost linear dependence of the charge neutrality point on $V_{BG}$ and $V_{TL}$, with a slope equal to the $C_{SiO_2}/C_{Al_2O_3}$ ratio.
Using $C_{SiO_2}=12$ nF/cm$^2$ for the bottom SiO$_2$ dielectric, we determine the inter-layer dielectric capacitance to be $C_{Al_2O_3}=340$ nF/cm$^2$.
The capacitance values are confirmed by Hall measurements.

\begin{figure}
\centering
\includegraphics[scale=0.4]{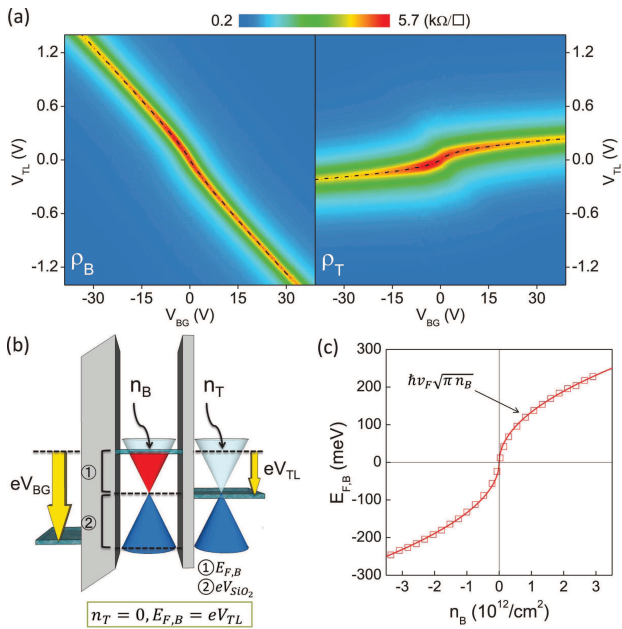}
\caption {\small{(color online) (a) Contour plots of $\rho_{B}$ (left panel) and $\rho_{T}$ (right panel) measured as a function of $V_{BG}$ and $V_{TL}$, at $T=0.4$ K.
The bottom layer responds to $V_{BG}$ and $V_{TL}$ similar to a dual-gated monolayer. (b) Band diagram of a graphene double layer with the top
layer at the charge neutrality point. The inter-layer bias is equal to the bottom layer Fermi energy. (c) Bottom layer Fermi energy vs. carrier
concentration, determined using data of panel (a). The symbols are experimental data, while the solid line represents the expected
$\hbar v_F \sqrt{\pi n_B}$ dependence.}}
\end{figure}

The top layer resistivity shows the characteristic ambipolar behavior as a function of $V_{TL}$, and with a weaker $V_{BG}$ dependence.
Let us examine more closely the top layer charge neutrality point dependence on $V_{BG}$ and $V_{TL}$. If we consider the top layer at
the charge neutrality point, setting $n_T=0$ in Eq. (2) yields:

\begin{equation}
eV_{TL}=E_F(n_B)
\end{equation}

This equation contains a simple, yet remarkable result. The inter-layer bias required to bring the top layer at the charge neutrality point is equal
to the Fermi energy of the opposite layer, in units of eV [Fig. 2(b)]. Consequently, tracking the top layer charge neutrality point in the
$V_{BG}$ - $V_{TL}$ plane [dash-dotted trace in Fig. 2(a) left panel], results in a measurement of the bottom layer Fermi energy as a function of $V_{BG}$.
Furthermore, setting $n_{T}=0$ in Eq. (1), and using Eq. (3) allows for $n_B$ to be determined as a function of $V_{BG}$ and $V_{TL}$ along the top layer
charge neutrality line of Fig. 2(a):

\begin{equation}
V_{BG}-V_{TL}=en_B/C_{SiO_2}
\end{equation}

Equations (3) and (4) provide a direct measurement of the bottom layer Fermi energy as a function of density. To illustrate this, in Fig. 2(c) we
show the bottom layer Fermi energy $E_{F,B}$ as a function of $n_{B}$, determined using Fig. 2(a) data and Eqs. (3) and (4). The $E_F$ values are in
excellent agreement with the $E_F(n_B)=\hbar v_F \sqrt{\pi n_B}$ dependence expected for the linear energy-momentum dispersion of graphene,
and with an extracted Fermi velocity of $v_F=1.15\times10^{8}$ cm/s.

In the following we show that the above method applies equally well to an electron system in high magnetic fields, allowing a direct
measurement of Landau level (LL) energies and broadening. In Fig. 3(a) we show the contour plots of $\rho_T$ (top panel) and $\rho_B$
(bottom panel) measured as a function of $V_{BG}$ and $V_{TL}$ in an applied perpendicular magnetic field $B=8$ T. Both layers show
quantum Hall states (QHS) marked by vanishing resistitvities at filling factors $\nu=4(N+\frac{1}{2})$, consistent with a graphene monolayer \cite{novoselov2005,zhang2005}. The integer $N$ represents the Landau level index. The top panel of Fig. 3(a) data shows a step-like dependence of the top layer charge neutrality point on $V_{BG}$ and $V_{TL}$. Similarly to Fig. 2, substituting $eV_{TL}$ with $E_{F,B}$ at the top layer charge neutrality line in Fig. 3(a) (top panel) provides a mapping of $E_{F,B}$ as a function of $V_{BG}$. To visualize this, the top layer charge neutrality line in the $V_{BG}-V_{TL}$
plane is superposed with the $\rho_{B}$ contour plot of Fig. 3(b) (bottom panel), which shows step-like increments of
$E_{F,B}$ coinciding with the QHSs of the bottom layer.

\begin{figure}
\centering
\includegraphics[scale=1.1]{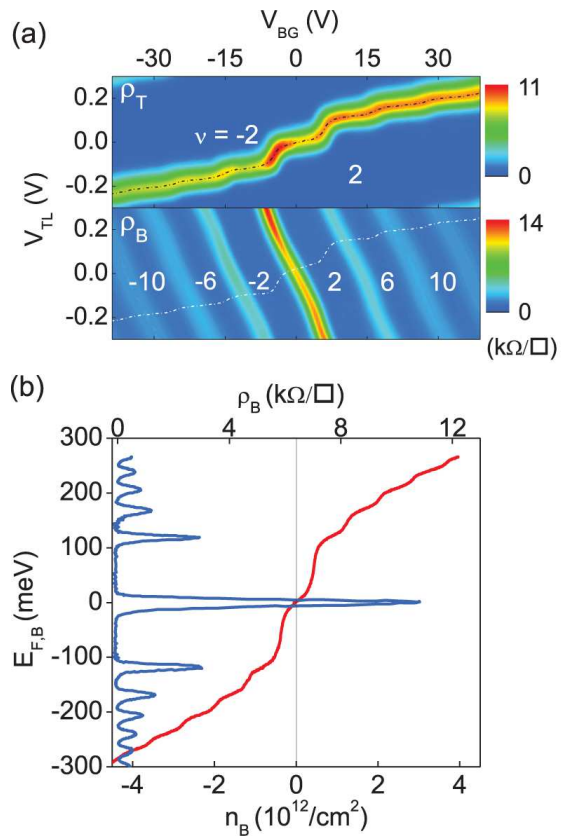}
\caption {\small{(color online) (a) $\rho_{T}$ (top) and $\rho_{B}$ (bottom) contour plots measured as a function of $V_{BG}$ and $V_{TL}$
at $B=8$ T, and $T=0.4$ K. Both layers show quantum Hall states marked by vanishing longitudinal resistance at filling factors $\nu=\pm 2, 6, 10$,
consistent with mono-layer graphene. The top layer charge neutrality line (dashed line) shows a step-like behavior, with the steps matching
the bottom layer quantum Hall states. (b) $\rho_{B}$ (blue line, top axis) vs. $E_{F,B}=eV_{TL}$, and $E_{F,B}$ vs. $n_{B}$
(red line, bottom axis) determined from the top layer charge neutrality line of panel (a). The $E_{F,B}$ values at the peak positions
of $\rho_{B}$ provide the Landau level energies.}}
\end{figure}

Figure 3(b) shows $E_{F,B}$ vs. $n_B$ at $B=8$ T determined by tracking the top layer charge neutrality line in the $V_{TL} - V_{BG}$ plane in Fig. 3(a),
and using Eqs. (3) and (4) to convert $V_{TL}$ and $V_{BG}$ into $E_{F,B}$ and $n_B$, respectively. In addition, Fig. 3(b)
shows $\rho_{B}$ vs. $E_{F,B}$, determined by tracking the bottom layer resistivity along the top layer charge neutrality
line [dashed-dotted line of Fig. 3(a)]. Figure 3(b) data manifestly shows the staircase-like behavior expected for the
Fermi level dependence on density for a two-dimensional electron system in a perpendicular magnetic field.
The peaks in the $\rho_{B}$ vs. $E_{F,B}$ data of Fig. 3(b), corresponding to the Fermi
level lying in the LL center and probing extended states, correlate with plateaus in the $E_{F,B}$ vs. $n_{B}$,
associated with the large LL density of states. The peaks in the $\rho_{B}$ vs. $E_{F,B}$ data of Fig. 3(b) provide
a direct measurement of the LL energy. Figure 4(a) summarizes the bottom graphene layer LL energy as a function of index
($N$) at $B=8$ T. The experimental data is in excellent agreement with the theoretical dependence $E_N=\pm v_F \sqrt{2 \hbar e B |N|}$,
corresponding to a Fermi velocity $v_F=1.17\times10^{8}$ cm$/$s, a value less than 2$\%$ different than the Fermi velocity
determined at $B=0$ T using Fig. 2 data.

In Figure 4(b) we compare the $E_{F,B}$ vs. $n_B$ data determined experimentally at $B=8$ T, with calculations.
Assuming a Lorentzian distribution of the disorder-induced LL broadening, the density of states $D(E)$ writes:
\begin{equation}
D(E)=\frac{4e}{h}B\sum_N{\frac{1}{\pi}\frac{\gamma_{N}}{(E-E_N)^2+\gamma_N^2}}
\end{equation}
with $\gamma_N$ being the broadening of the $N$-th LL. The carrier density ($n$) dependence on $E_F$ in the limit $T=0$ K is:
\begin{equation}
n(E_F)=\int_0^{E_F}D(E)dE
\end{equation}
Using Eqs. (5) and (6), the best fit to $E_{F,B}$ vs. $n_B$ data is obtained for $\gamma_0=14$ meV, and $\gamma_N=6.5$ meV
for $|N|>0$. The summation in Eq. (5) does not converge if carried out to infinity, and a high-energy cut-off is customarily used.
For the calculations of Fig. 4(b) we used $|N|\leq100$ in Eq. (5), corresponding to a 1 eV cut-off energy; increasing
the cut-off LL index to 1,000, will change the best fit $\gamma$ value by less than 0.5 meV.
The lower inset of Fig. 4(b) shows a comparison of the $E_{F,B}$ vs. $n_B$ experimental data with calculations
using the same broadening for the $N=0$ LL as the upper and lower LLs, $\gamma=6.5$ meV.
The larger broadening of the $N=0$ LL by comparison to the other LLs is an interesting finding.
A theoretical study \cite{zhu2009}, which examined the impact of static disorder on LL broadening in graphene without considering interaction
showed that the $N=0$ LL broadening is the same as for the other LLs. On the other hand electron-electron
interaction can impact the broadening of the four-fold degenerate $N=0$ LL, and experimental data
on exfoliated graphene on SiO$_2$ substrates show a splitting of the $N=0$ LL in high, $B=45$ T magnetic fields \cite{zhang2006},
explained as a many-body effect. Lastly, we note that a Gaussian-shaped Landau levels density of states yields worse fits to Fig. 4 data,
by comparison to the Lorentzian shape density of states. Scanning tunneling microscopy studies \cite{li,miller}, and
compressibility measurements in graphene \cite{jp2010} also favor the Lorentzian LL lineshape by comparison to the Gaussian one.
A recent theoretical study argues that LL local density of states has a Lorentzian lineshape while
the total density of states is Gaussian \cite{zhu2011}. Presumably, the sample size examined here, defined by a 4 $\mu$m Hall bar width
coupled with the 8 $\mu$m top layer contact spacing is sufficiently small such that the Lorentzian LL line-shape dominates.

\begin{figure}
\centering
\includegraphics[scale=0.3]{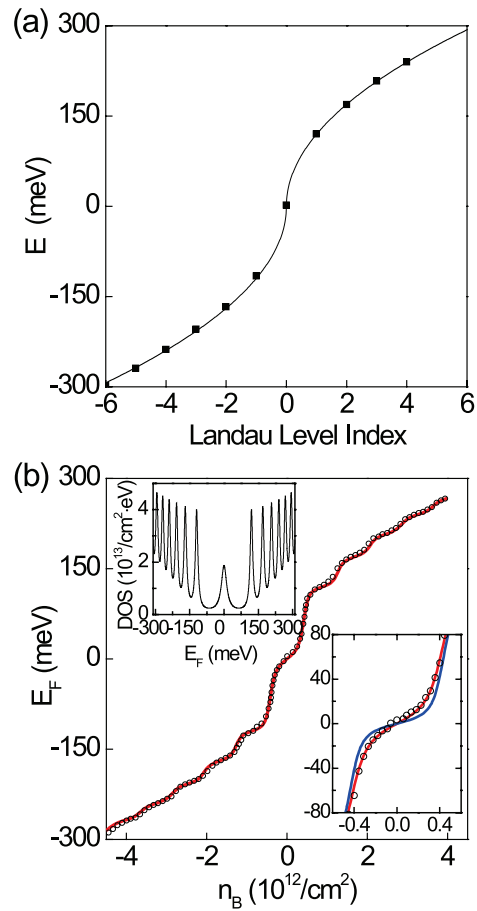}
\caption {\small{(color online) (a) Landau level energy in mono-layer as a function of index ($N$). The symbols are experimental data determined
from the $E_{F,B}$ positions at the $\rho_{B}$ peaks in Fig. 3(b). The solid line is the theoretical $\pm v_F \sqrt{2 \hbar e B |N|}$
dependence corresponding to $v_{F}=1.17\times10^{8}$ cm$/$s. (b) Bottom layer Fermi energy vs. density at $B=8$ T. The symbols represent
experimental data, and the solid (red) line is a fit assuming a Landau level Lorentzian line shape. The best fit is
obtained for a LL broadening of $\gamma_N=6.5$ meV for $|N|>0$, and $\gamma_0=14$ meV. The upper inset shows the calculated density of states
corresponding to the best fit to experimental data. The lower inset shows the $E_F$ vs. $n_B$ data in the vicinity of zero density. The symbols represent
experimental data, and the lines are calculations assuming $\gamma_0=6.5$ meV (blue line) and $\gamma_0=14$ meV (red line).}}
\end{figure}

In summary, we present a method to determine the Fermi energy in a two-dimensional electron system, using a double
layer device structure with one layer consisting of graphene. We illustrate this technique by probing the Fermi energy in
a separate graphene layer as a function of density at zero, and in a high magnetic field, and determine with high accuracy the
Fermi velocity, and the Landau level broadening. The technique sensitivity is independent of the sample dimensions,
which makes it applicable to a variety of nanoscale materials.

We thank C. P. Morath and M. P. Lilly for technical discussions. This work was supported by NRI, ONR, and Intel.
Part of this work was performed at the National High Magnetic Field Laboratory,
which is supported by NSF (DMR-0654118), the State of Florida, and the DOE.

\newpage


\begin{thebibliography}{10}
\bibitem{gate_bias}
For simplicity $V_{BG}$ and $V_{TL}$ are referenced with respect the bias values
at which both layers are at the Dirac point: $V_{BG}=8$ V, $V_{TL}=-0.01$ V.
\end{thebibliography}
\end{document}